# Semi-empirical versus Theoretical Stellar Population Models: a comparison with Star Clusters

Randa Asa'd,[1] Paula R. T. Coelho,[2] Johina M. John,[1] Igor Chilingarian,[3,4] Gustavo Bruzual,[5] and Stephane Charlot[6]

[1]*American University of Sharjah, Physics Department, P.O. Box 26666, Sharjah, UAE*
[2]*Universidade de São Paulo, Instituto de Astronomia, Geofísica e Ciencias Atmosféricas, Rua do Matão 1226, 05508-090 São Paulo, Brazil*
[3]*Center for Astrophysics – Harvard and Smithsonian, 60 Garden St. MS09, Cambridge, MA, 02138, USA*
[4]*Sternberg Astronomical Institute, M.V. Lomonosov Moscow State University, 13 Universitetsky prospect, Moscow, 119991, Russia*
[5]*Instituto de Radioastronomía y Astrofísica, UNAM, Campus Morelia, Michoacán, C.P. 58089, México*
[6]*Sorbonne Universite, CNRS, UMR7095, Institut d'Astrophysique de Paris, F-75014 Paris, France*

## ABSTRACT

Stellar population synthesis (SPS) models are a key tool for deriving the age, metallicity, radial velocity and reddening of star clusters from their integrated spectra. Using a sample of 129 star clusters with high-quality spectra, we analyze the uncertainties associated with selecting an empirical versus a theoretical stellar spectral library in the SPS models. We find that the fits from the different models agree on the goodness of fit metrics and inferred reddening. However, the derived age and metallicity can be affected by the choice of the stellar library, with synthetic libraries tending to give lower age and metallicity, especially for spectra with low SNR. Ages and reddening values from SSP-equivalent fits are consistent with the multi-population fits, however, SSP-equivalent metallicities are affected by the coarse coverage of the SPS grid in [Fe/H]. When comparing the spectral fitting results with the literature, we find that (1) all models underestimate age for old and metal-poor systems; (2) on average, SPS models based on synthetic stellar libraries better match the isochrone ages and metallicities from high-resolution stellar spectroscopy.

*Keywords:* stars: atmospheres – galaxies: star clusters: general.

## 1. INTRODUCTION

Our understanding of the universe in the past decades has evolved significantly thanks mainly to three factors: a) the advanced observational tools (e.g., Sánchez et al. 2012); b) the availability of detailed stellar population synthesis (SPS) models (e.g., Bruzual & Charlot 2003; Schaye et al. 2015; Vazdekis et al. 2010a); and c) the development of computational programs designed to compare accurate observations with sophisticated models (e.g., Cid Fernandes et al. 2005a; Cappellari 2017a).

For the majority of the unresolved extra-galactic star clusters, we must rely on either integrated photometry or spectra to derive the age and metallicity of the stellar population (e.g., Ahumada et al. 2002; Puzia et al. 2005, 2006; Santos et al. 2006; Palma et al. 2008; Talavera et al. 2010; Cid Fernandes & Gonzalez Delgado 2010; Asa'd et al. 2013; Asa'd 2014; Chilingarian & Asa'd 2018).

Corresponding author: Randa Asa'd
raasad@aus.edu

A method to interpret integrated light observations in ample use by the community is the so-called spectral fitting technique: an observed spectrum is compared to a collection of spectral templates pixel-by-pixel in wavelength, until a "best match" is achieved via a pre-chosen minimization algorithm (e.g. Chilingarian et al. 2007; Koleva et al. 2009; Cappellari 2017b; Gomes & Papaderos 2017; Wilkinson et al. 2017; Johnson et al. 2021). A common practice is to use simple stellar population (SSP) spectra (e.g., Bruzual & Charlot 2003; Le Borgne et al. 2004; Sánchez-Blázquez et al. 2006; Verro et al. 2022) as templates in the fitting codes. The ability of spectral fitting to derive reliable stellar population parameters, such as age, metallicity, and reddening of star clusters, has been investigated in several works in the literature (e.g. Koleva et al. 2008; Walcher et al. 2009; Cid Fernandes & González Delgado 2010; González Delgado & Cid Fernandes 2010; Dias et al. 2010; Cezario et al. 2013; Gonçalves et al. 2020; Goudfrooij & Asa'd 2021).

Using a sample of 21,000 mock star clusters, Asa'd & Goudfrooij (2020) show that the precision achieved by full-spectral fitting methods depends on both the signal-to-noise ratio (SNR) and the wavelength range covered by the fitted spectral energy distribution (SED). They conclude that for SNR ≥ 50, spectral fitting recovers the age of the mock stellar population (SSPs in their case) with an overall precision of 0.1 dex for age in the ranges $7.0 \leq \log(\text{age/yr}) \leq 8.3$ and $8.9 \leq \log(\text{age/yr}) \leq 9.4$. For the age ranges $8.3 \leq \log(\text{age/yr}) \leq 8.9$ and $\log(\text{age/yr}) \geq 9.5$, with significant contributions from asymptotic giant branch and red giant branch stars, respectively, the age uncertainty rises to about ±0.3 dex. The precision of the full-spectrum fitting method in deriving possible age spreads within a star cluster was examined by Asa'd et al. (2021) using 118,800 mock star cluster populations. They find that the mean age derived for the younger populations generally matches the actual age to within 0.1 dex up to $\log(\text{age/yr}) \leq 9.5$. The precision decreases for $\log(\text{age/yr}) > 9.6$ for any mass fraction or SNR due to the similarity of the SEDs at these ages. In this series of studies they perform a systematic analysis comparing the age of star clusters obtained from isochrone fitting to colour-magnitude diagrams (CMDs) and integrated-light spectroscopy, using models based on the same sets of isochrones. Chilingarian & Asa'd (2018) use the integrated optical spectra of 15 star clusters with age from 40 Myr to 3.5 Gyr, finding good agreement between the age derived from CMDs and integrated spectra.

We remark that the stellar population models used in spectral fitting have uncertainties that may affect the values of the physical properties derived from the fits to observed spectra. The dependence of the inferred parameters on the choice of the SPS models has been noted both in (relatively) simple populations (e.g., SMC clusters in Dias et al. 2010) and in composite populations (e.g., galaxies in Coelho et al. 2009a). González Delgado & Cid Fernandes (2010) derive the age, extinction, and metallicity for a sample of star clusters using different SPS models. They find that the inferred ages are consistent within models and with values in the literature. However, there is little agreement between the metallicity and extinction obtained from different SPS models. These uncertainties have been assessed from different perspectives in the literature (e.g., Charlot et al. 1996; Yi et al. 1997; Yi 2003; Anders et al. 2004; Salaris & Cassisi 2007; Deng & Xin 2007; Conroy et al. 2009, 2010; Conroy & Gunn 2010; Riffel et al. 2011; Gonzalez-Perez et al. 2014; Magris C. et al. 2015)

It is not straightforward to identify the origin of these differences. One source of uncertainty is the choice of an empirical versus a theoretical stellar spectral library as a basic ingredient in the SPS models. On the one hand, theoretical libraries (e.g., Leitherer et al. 1999; Gonzalez Delgado et al. 2005; Coelho et al. 2007a; Maraston 2005; Percival et al. 2009; Leitherer et al. 2014) provide high resolution stellar spectra for a wide wavelength range with well-defined atmospheric parameters and infinite signal-to-noise ratio (SNR). However, our limited knowledge of the exact physics of stellar atmospheres restricts our ability to accurately reproduce observations (Bessell et al. 1998; Kučinskas et al. 2005; Kurucz 2006; Bertone et al. 2008; Coelho et al. 2009b; Plez 2011; Lebzelter et al. 2012; Sansom et al. 2013; Knowles et al. 2019). On the other hand, SPS models that use empirical stellar libraries (e.g., Vazdekis et al. 2010b, 2016) reproduce quite well the observed spectral features but lack the coverage of stellar physical parameters provided by the theoretical libraries. By construction, empirical libraries are biased towards stellar systems where we can resolve individual stars, which makes it impossible to fully sample the parameter space required to study a large diversity of stellar populations. A classic example is the need for theoretical predictions to model $\alpha$-enhanced systems (e.g., Thomas et al. 2003; Coelho et al. 2007b; Walcher et al. 2009; Percival et al. 2009). Likewise, empirical libraries in general cover narrower wavelength ranges than their theoretical counterparts.

A valid question is then, which kind of library should be favoured when using SPS models? Coelho et al. (2020, hereafter CBC20) investigated this question. SPS models were created using the same set of evolutionary tracks for three choices of the stellar spectral library: *(a)* the empirical MILES library (Sánchez-Blázquez et al. 2006; Cenarro et al. 2007), *(b)* a theoretical spectral library mimicking the coverage in stellar parameters of the MILES library, named SynCoMiL by the authors, and *(c)* an extended version of the theoretical library of Coelho (2014), hereafter the Coelho14ext library. CBC20 performed spectral fits to a sample of nearby galaxies using the three sets of SPS models. They conclude that the models based on the MILES and SynCoMiL libraries may underestimate the age of the stellar population due to their limited coverage of the HR diagram (HRD). Conversely, the SPS models based on the Coelho14ext theoretical library may underestimate the metallicity of the integrated stellar population. Their conclusions are based on galaxies, systems for which we do not have a reference or "ground truth" value. Re-assessing this question using star clusters – systems for which we can obtain the age and metallicity from resolved population studies – is thus timely.

This work addresses this gap. We perform spectral fits to the integrated spectra of star clusters covering a wide range in age and metallicity using the CBC20 SPS models.

Section 2 describes the observational data used in our analysis. Section 3 describes the SPS models and the fitting code employed. The results are presented in Section 4. The discussion and conclusions follow in Sections 5 and 6, respectively.

## 2. INTEGRATED SPECTRA OF STAR CLUSTERS

We analyze 129 star clusters in the Milky Way, LMC, and SMC. Figure 1 illustrates the age-metallicity relation of our sample; the parameters are based on the literature values compiled by Gonçalves et al. (2020).

### 2.1. *SOAR DATA*

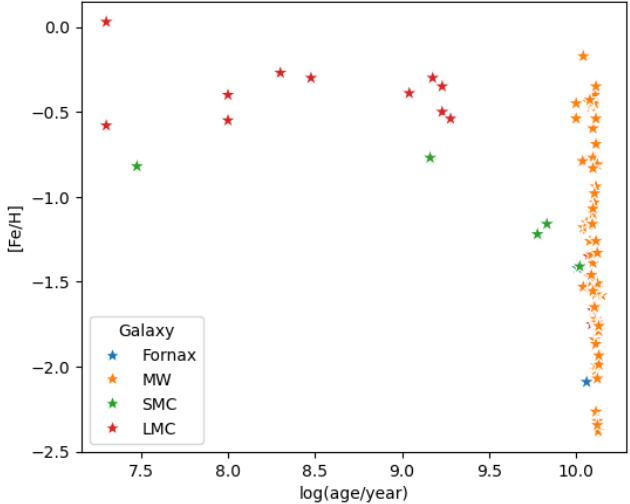

**Figure 1**: Age and metallicity of the star clusters used in this work. Symbols are color-coded according to host galaxy.

A sample of 44 integrated stellar spectra was obtained in two different observing runs on the 4-m SOAR telescope (Clemens et al. 2004) with the Goodman Spectrograph. During the first run (December 2011, Asa'd et al. 2013) 11 clusters were observed in the long-slit low-resolution mode (1.03 arcsec wide slit; 600 gpm VPH grating; R=1500; $3600 < \lambda < 6250$ Å). The second data set was observed with the long-slit 1 arcsec wide slit; 930 gpm VPH grating; R=2100; $3710 < \lambda < 5430$ Å) in January 2018. Our targets span the age range from 10 Myr to 3 Gyr. These observations will be presented in detail in Asad et al. (in prep.).

To collect an integrated spectrum, we scanned a cluster across the slit using non-sidereal tracking. The data reduction is as described in Chilingarian & Asa'd (2018). For the SOAR data, we started with bias subtraction and flat fielding, then subtracted the sky background and performed relative flux calibration using a standard star. We integrated the spectrum along the slit, creating the final one-dimensional data product.

## 2.2. *WAGGS DATA*

To complement our sample towards older age and lower metallicity clusters, we include 85 star clusters from the Wide Field Spectrograph (WiFeS) Atlas of Galactic Globular cluster Spectra (WAGGS), a comprehensive library of Milky Way and Local Group globular cluster integrated spectra (Usher et al. 2017). WiFeS (Dopita et al. 2007) is a dual-beam image-slicing integral field spectrograph offering a broad 25×38 arcsec field of view, with 1.0 arcsec square spaxels, effectively covering the entire field.

The WiFeS spectra were acquired in the Mount Stromlo and Siding Spring Observatory 2.3m telescope. Further insights into the on-telescope performance of WiFeS can be found in the comprehensive studies by Dopita et al. (2007, 2010).

## 3. MODELS AND METHOD

We use the CBC20 SPS models. These models were built using the Bruzual & Charlot (2003) GALAXEV code with the PARSEC (Bressan et al. 2012; Chen et al. 2015) stellar evolutionary tracks for metallicity Z = 0.0002, 0.004, 0.008, 0.017 and 0.03, assuming the Chabrier (2003) initial mass function. The CBC20 models are available for three distinct stellar spectral libraries:

1. The empirical MILES library (Sánchez-Blázquez et al. 2006; Cenarro et al. 2007);

2. The theoretical SynCoMiL library, computed by CBC20, which mimics the MILES stellar library in terms of HRD coverage and wavelength resolution and sampling; and

3. The theoretical Coelho14ext library, a version of the Coelho (2014) library extended to cover chemical abundances considered by CBC20 (see their Table 3).

We use the STARLIGHT spectral fitting code (Cid Fernandes et al. 2005b) to infer the stellar population parameters of each cluster in our sample for each set of SPS models, as in Section 4.2 of CBC20. Two metrics are used to quantify the quality of the fit,

$$\chi^2 = \frac{1}{N} \sum_\lambda \frac{[(OF)_\lambda - (MF)_\lambda]^2}{\sigma_\lambda^2}, \quad (1)$$

and

$$\mathtt{adev} = \frac{1}{N} \sum_\lambda \frac{|(OF)_\lambda - (MF)_\lambda|}{(OF)_\lambda}, \quad (2)$$

the average relative deviation. $OF$ is the observed flux, $MF$ the model flux, $N$ the number of wavelength points in the spectrum, and $\sigma_\lambda = OF_\lambda/\mathrm{median}(SNR_\lambda)$. We fit the cluster spectra in the wavelength interval 3694 – 5396 Å, common to the different datasets, adopting a 5$\sigma$ clipping. The spectra of the second SOAR and WAGGS samples have higher spectral resolution than the SPS models. They were convolved with a Gaussian of wavelength dependent width to degrade their spectral resolution to that of the models (FWHM = 2.3 Å). We select the extinction law according to the cluster host galaxy: Gordon et al. (2003) for the SMC and LMC, and Cardelli et al. (1989) for the Milky Way.

Figure 2 shows the fit to the spectrum of NGC2058 for the different SPS models. The fits for the other clusters and the table containing the derived stellar population parameters are available as supplementary material.

We use the same definitions of inferred parameters as Gonçalves et al. (2020). For age, we define:

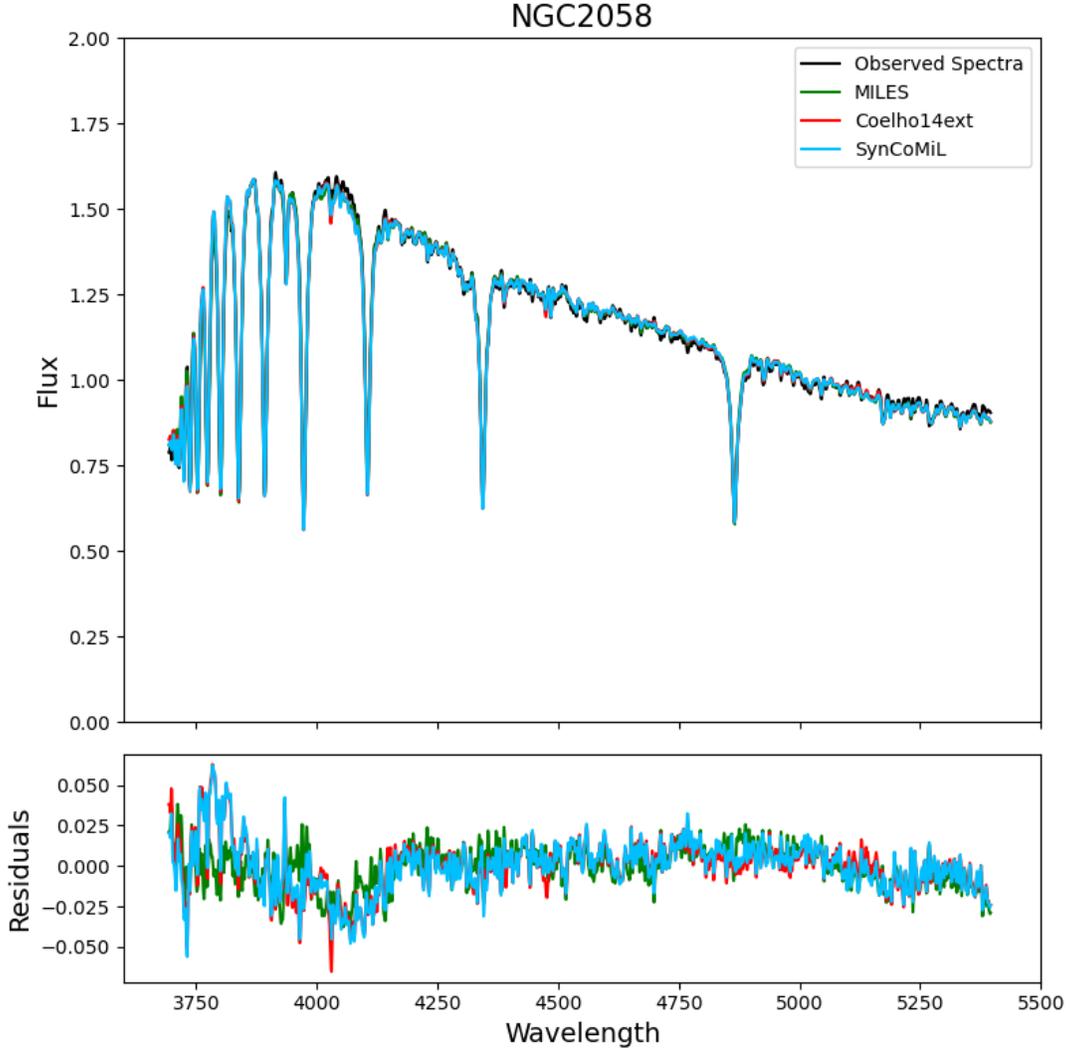

**Figure 2**: Spectral fit to the SED of NGC2058 using CBC20 SPS models for the MILES, SynCoMiL and Coelho14ext stellar libraries. The observed spectrum is show in *black* and the fitted spectra in *green*, *cyan* and *red*, respectively. The residuals (fitted-observed) are displayed in the bottom panel.

*Luminosity-weighted logarithmic age*—hereafter light-weighted age (LWA),

$$LWA = \langle \log(age) \rangle_L = \frac{\sum_j x_j \cdot \log(age)_j}{\sum_j x_j}, \quad (3)$$

is the average luminosity-weighted logarithmic age of a composite stellar population, determined by weighting the logarithmic age of the $j^{th}$ base element (SSP) by its contribution $x_j \cdot L(\lambda_o)$ to the total luminosity at the normalization wavelength $\lambda_o$. Since by construction for all base elements $L(\lambda_o) = 1$, $L(\lambda_o)$ cancels out in Eq. (3). In our case $\lambda_o = 5000$ Å. For clarity, from here on $\langle \log(age) \rangle_L$ will be denoted as log(age). See also González Delgado et al. (2021).

*SSP-equivalent age (SSP-EA)*—Age of the SSP in our spectral basis that best fits the observed spectrum (the one with the smallest $\chi^2$).

Equivalent definitions of LWZ = $\langle \log(Z) \rangle_L$ and SSP-EZ are used in the case of metallicity Z or [Fe/H].

4. RESULTS

In the left column panels of Figure 3 we compare the results obtained with the MILES and SynCoMiL CBC20 models, illustrating the *synthetic effect*, i.e., the consequences of replacing in the SPS models an empirical by a synthetic library at a fixed HRD coverage. The right panel columns compare the results from the models based on the Coelho14ext and SynCoMiL libraries, illustrating the *coverage effect*, i.e., the consequences of bringing the limited HRD coverage of the SynCoMiL and MILES libraries to an almost complete HRD coverage. Dots are color coded according to the goodness-of-fit. The *blue* and *green* dots represent results from fits with $\chi^2 \leq 3$ and $\chi^2 > 3$ for both SPS models, respectively. The *black* dots correspond to fits with $\chi^2 \leq 3$ for one SPS model and > 3 for the other.

The top panels display the resulting adev, Eq. (2).

The five outlier clusters (NGC1846, NGC2004, NGC6440, NGC6528 and NGC6553) are shown as *red* dots. The figure shows the consistency of the cluster properties derived from SPS models based on purely theoretical and purely empirical stellar libraries. The major discrepancy occurs for [Fe/H] (bottom right panel), where the Coelho14ext models predict lower values than their SynCoMiL counterpart.

The error bars for the SOAR sample in Figure 3 are adopted from Cid Fernandes & González Delgado (2010) and González Delgado & Cid Fernandes (2010), who discuss the precision of the STARLIGHT full-spectrum fitting algorithm. For clusters with log (age/year) > 9, we find an uncertainty of log (age/year) = 0.4, and log (age/year) = 0.2 for clusters with log (age/year) < 9, shown in *purple* in the figure.

Similarly, for [Fe/H] we obtain an uncertainty of 0.25 for clusters with [Fe/H] > -0.75 and 0.5 for clusters with [Fe/H] < -0.75. For clusters with derived $A_v > 0.4$ mag, we adopt an uncertainty of 0.3 mag and of 0.1 mag for clusters with $A_v < 0.4$ mag. For the WAGGS clusters, the uncertainty in the inferred stellar parameters is adopted from Table 4 of Gonçalves et al. (2020).

In general, the $\chi^2$ values obtained with the SPS models based on theoretical libraries are larger than for the empirical MILES library, in agreement with what was reported by CBC20. The inferred $A_v$ is consistent among the three SPS models, independently of the SNR.

The values of log (age/year) derived from the theoretical SynCoMiL library SPS models are slightly lower than those derived from SPS models based on the empirical MILES library. For log (age/year) < 8.5, there is a slight spread in the values inferred from SPS models based on the SynCoMiL library compared to the ones using the Coelho14ext library. This spread is associated with higher values of $\chi^2$. The derived [Fe/H] is slightly lower using SPS models based on the theoretical SynCoMiL library than those based on the empirical MILES library. The largest spread is seen when results obtained with SPS models based on the SynCoMiL library are compared to those derived with the Coelho14ext library, mainly for fits with $\chi^2 > 3$.

We repeat our analysis now using a single SSP model fit instead of the multi-population fits.

The results are presented in Figure 4. The upper panels of Figure 4 show that the outliers in the adev comparison are the same as before. A comparison of the lower panels of Figures 3 and 4 show that the multi-population fits are less dependent on the stellar library used in the SPS models than the single SSP fits.

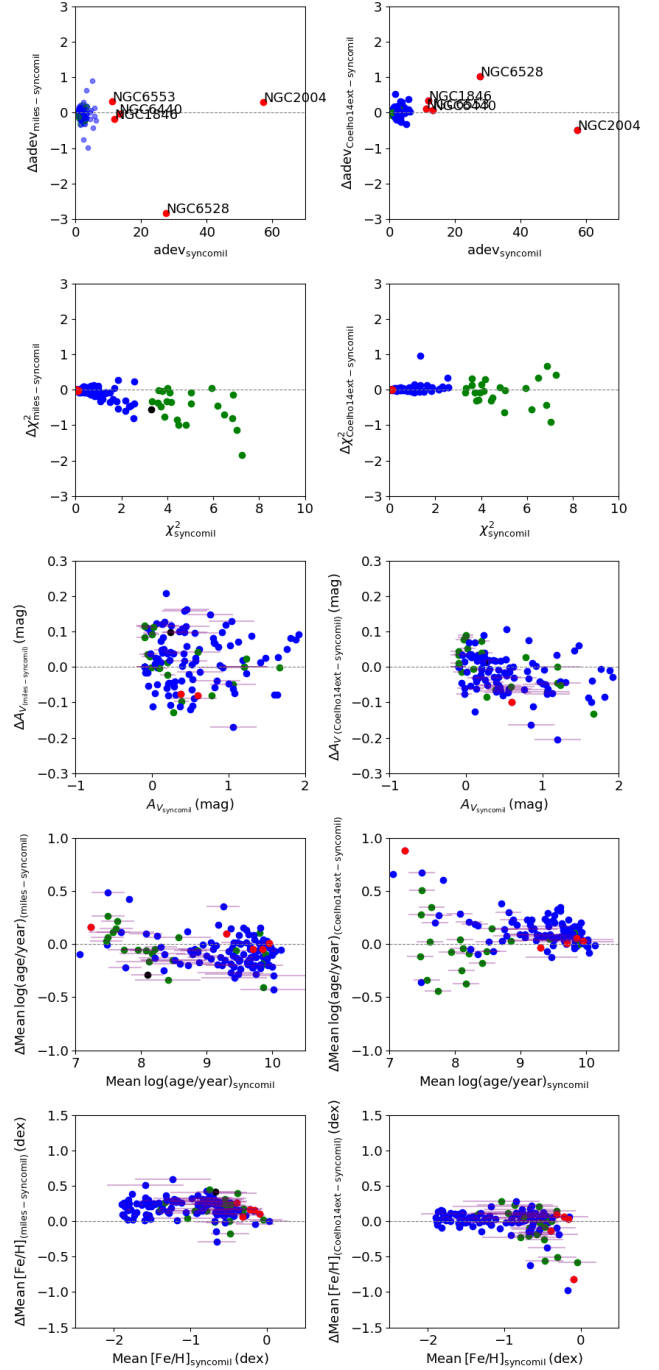

**Figure 3**: Comparison of results for adev, $\chi^2$, log(LWA), light-weighted [Fe/H] and $A_v$ obtained from multi-population fits. *Left column:* Difference in derived values between MILES and SynCoMiL SPS models. *Right column:* Difference in derived values between Coelho14ext and SynCoMiL SPS models. Dots are color coded according to the goodness-of-fit: *blue* and *green* dots represent fits with $\chi^2 \leq 3$ and $\chi^2 > 3$ for both SPS models, respectively. The *black* dots correspond to fits with $\chi^2 \leq 3$ for one SPS model and > 3 for the other.

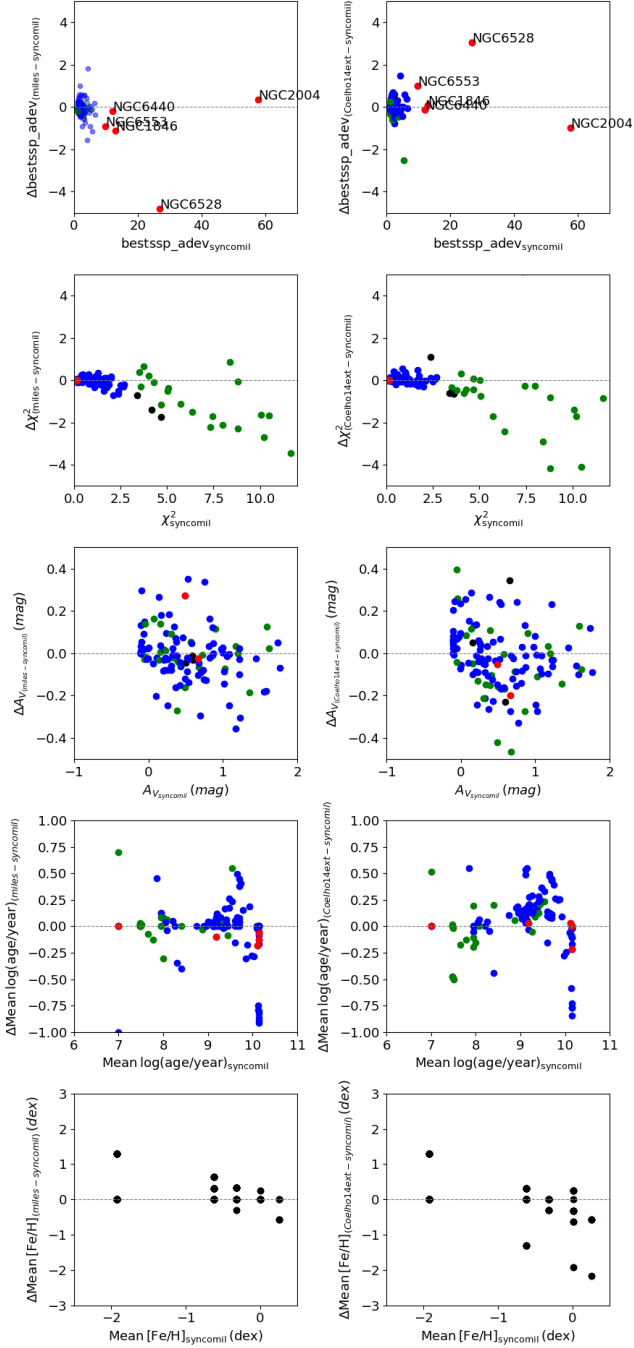

**Figure 4**: Same as Figure 3 but for parameters obtained using single SSP fits. Due to the overlap of points of different colors in the [Fe/H] panel, all dots are shown in *black*.

The lower panel of Figure 4 shows the well-known "quantization" issue on metallicity when performing single SSP fits, which arises due to the coarse metallicity grid of the adopted PARSEC stellar library. In Figure 5 we compare the light-weighted age and $A_v$ from the multi-population fits with the single SSP results. In general, the light-weighted age is larger than the single SSP value, and the light-weighted $A_v$ tends to be lower than its single SSP counterpart.

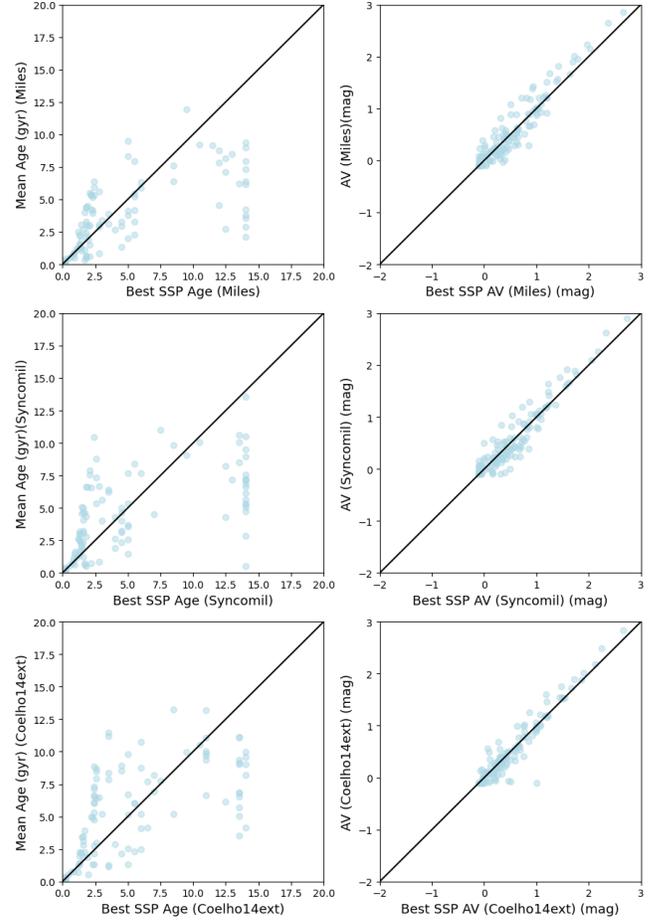

**Figure 5**: Comparison of light-weighted age and $A_v$ from the multi-population fits with the age and $A_v$ from the single SSP fits.

## 5. DISCUSSION

### 5.1. *Comparison with Coelho et al. (2020)*

Figure 3 illustrates the outcome of our spectral fits in the same manner as Figure 12 of CBC20 (based on the spectral fits to galaxy SEDs). The y-axis in each panel shows the difference in the derived values between the libraries. The *left* column of Figure 3 compares the results from MILES vs. SynCoMiL CBC20 SPS models, illustrating the synthetic effect: changes introduced by replacing an empirical stellar library by a synthetic one with the same HRD coverage. The *right* column compares results from Coelho14ext vs. SynCoMiL CBC20 SPS models, illustrating the coverage effect: changes introduced by increasing the HRD coverage using synthetic libraries. In what follows we discuss our results in detail.

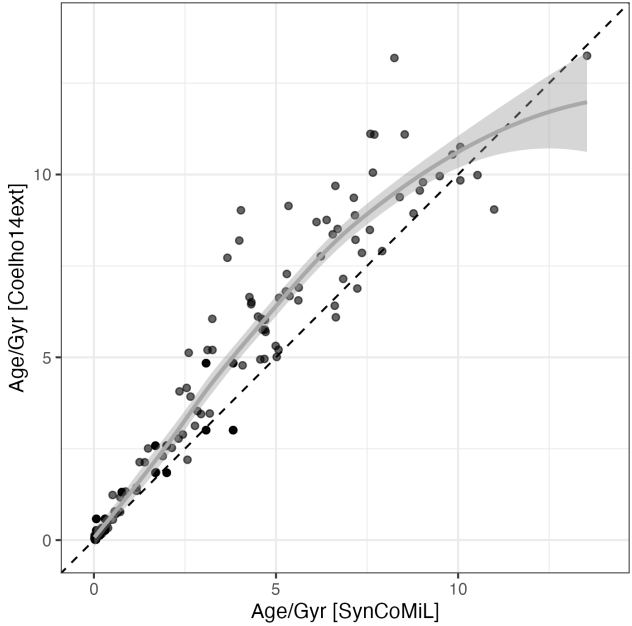

**Figure 6**: Comparison between the age derived using SPS models based on the Coelho14ext (y-axis) vs the SynCoMiL (x-axis) stellar libraries. Age is shown in a linear scale to highlight the age interval where the comparison deviates more from the 1-to-1 line. The grey line represents the LOESS (locally estimated scatterplot smoothing) fit to the data, and the shaded area shows its confidence interval.

#### 5.1.1. *The synthetic effect*

The results for adev, $\chi^2$ and $A_V$ are qualitatively similar to those in CBC20, namely, the synthetic effect increases slightly the adev and $\chi^2$ metrics and $A_V$ remains virtually unaffected. In the case of age, our results show a more pronounced systematic difference than in CBC20. Ages based on MILES models are lower than ages based on SynCoMiL models, albeit the difference is within the adopted error bars. Regarding metallicity, our analysis covers a larger interval than in CBC20. We see a distinct systematic difference in that fits based on MILES models return a higher metallicity than those based on SynCoMiL models. Our results do not reproduce the pattern that this difference is more pronounced at low metallicity noted by CBC20. Yet, we remark that our sample does not contain many clusters around solar metallicity, as is the case of the galaxy sample in CBC20. We note that the combined effect of lower age and higher metallicity is qualitatively consistent with the well known age-metallicity degeneracy present in stellar populations.

#### 5.1.2. *The coverage effect*

The effects on adev and $A_V$ are similar to those reported by CBC20. These quantities are virtually unaffected by using one or the other synthetic stellar library in the SPS models. The impact on $\chi^2$ is distinct: for our star cluster sample this metric is unaffected by the coverage effect, while CBC20 find a hint of slightly larger $\chi^2$ for Coelho14ext based models. CBC20 report that intermediate age estimates are larger using Coelho14ext based SPS models. This is not immediately visible in Figure 3, but it can be seen if we plot our results in a linear rather than in a logarithmic scale (see Figure 6). We thus confirm the result from the literature in the sense that the coverage effect is more relevant for intermediate age populations.

Regarding metallicity, we confirm the results in the literature in the sense that, for most of our sample, the derived metallicity is unaffected by the HRD coverage of the stellar library used in the SPS models. Yet, we highlight some clusters where [Fe/H] derived from Coelho14ext based SPS models is significantly lower (reaching [Fe/H] ~ -2) than when derived with SynCoMiL based models. These findings are puzzling but hinted in CBC20, where the metallicity comparison shows more outliers than other quantities. The most affected clusters are NGC2100, NGC1850 and NGC0330. Visual inspection of the spectral fits for these clusters do not reveal a pattern, and the reason for these clusters to be outliers has yet to be clarified. We note that although a large number of the outliers in Figure 3 are green dots, i.e., SED fits with $\chi^2 > 3$, there are also blue dot outliers, representing fits with $\chi^2 < 3$. Similarly, not all clusters fitted with $\chi^2 > 3$ are outliers in this figure.

### 5.2. *Results from a different full-spectrum fitting code*

We repeated our analysis using the Asa'd (2014); Asa'd & Goudfrooij (2020); Asa'd et al. (2021) full spectrum fitting tool (hereafter ASAD). The code uses SPS models from Vazdekis et al. (2010b) and returns the single SSP that best fits the data. The results are shown in Figure 7. We use Table 2 of Asa'd & Goudfrooij (2020) to estimate the error in the ASAD age estimate as a function of SNR. The results from ASAD are consistent with those from STARLIGHT for single SSP fits.

### 5.3. *Comparison with values from the literature*

A significant advantage of using star clusters is that we can directly compare the results from spectral fitting with ages and metallicities derived from resolved population studies. Thus, we have access to reference values which are independent of SPS models, namely ages derived from isochrone fitting to high-quality CMDs, and metallicities derived from high-resolution stellar spectroscopy (e.g. Gonçalves et al. 2020). For reference values we adopt the compilation of cluster ages and metallicities in Gonçalves et al. (2020, Table 1), supplemented with ages from Milone et al. (2023). In total, we have literature values for 27 star clusters in the SOAR sample and 84 clusters in the WAGGS sample. There are 5 star clusters in common to both samples:

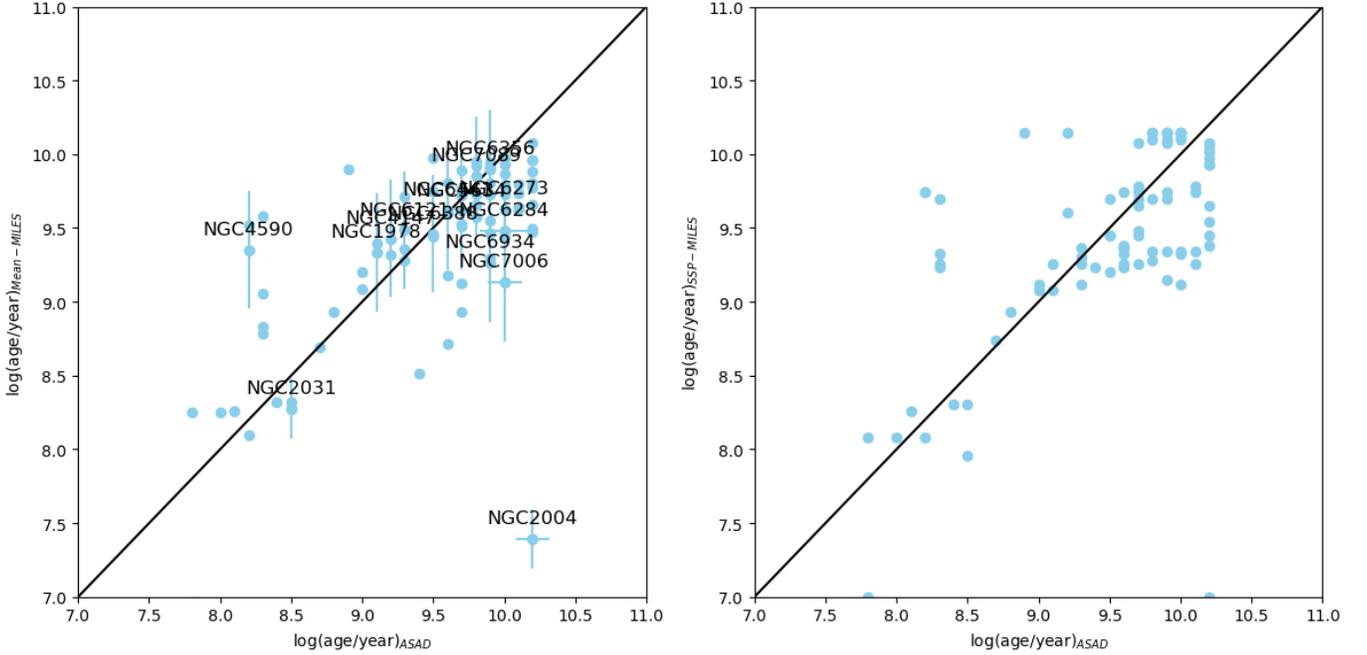

**Figure 7**: Comparison between the age inferred by STARLIGHT (y-axis) and ASAD (x-axis). In the *left* panel the vertical axis shows the light weighted age from a multi-population fit, whereas in the *right* panel and the horizontal axis of both panels, we show the age of the best fitting single SSP

NGC 0330, NGC 0416, NGC 1846, NGC 1850 and NGC 1978. We should note that a limitation of this work is that the dataset used is heterogeneous and the isochrone ages and the SSPs were not obtained with self-consistent stellar model libraries.

#### 5.3.1. *Ages*

For the star clusters in the LMC and SMC we compare the results of our fits with values from Milone et al. (2023) in Figure 8. We adopt a typical error in age of 20%. In Milone et al. (2023) the clusters with an extended main sequence turnoff (eMSTO) are assigned two different ages, based on fitting isochrones to the lower and upper part of the eMSTO. For our comparison, in those cases we use the average of these two values as the cluster age. Ages derived from STARLIGHT are in good agreement with literature, although we note that the SSP-equivalent ages tend to be systematically younger than literature.

In Figure 9 we compare the behavior of the ages for our entire sample. The top panel shows the $\Delta t$ (STARLIGHT minus literature) as a function of literature age. The bottom panel shows the density distribution of the $\Delta t$. There is a strong effect at the oldest ages, with a large spread and most clusters showing STARLIGHT ages significantly younger than the literature. By looking in the bottom panel, we see that MILES-based models are more strongly affected than the other SPS models. We discuss this effect later in the section.

#### 5.3.2. *Metallicities*

In Figure 10 we compare the metallicity derived from STARLIGHT spectral fits with that measured in high-resolution stellar spectra from the literature. The top panel shows the correlation of the two metallicities. The color-coded lines are linear fits to the results obtained for each set of SPS templates. The linear fits are remarkably parallel to each other, implying that the choice of SPS templates introduces at most an offset in the results. The linear fits are not parallel to the 1-to-1 line (gray solid line). The origin of this pattern is currently unclear. The bottom panel of Figure 10 shows the density distribution of $\Delta$[Fe/H] = [Fe/H]$_{\text{STARLIGHT}}$ - [Fe/H]$_{\text{LITERATURE}}$. Comparing the distributions associated to the SynCoMiL and MILES templates, we see another evidence that replacing an empirical library by a synthetic one (at fixed HRD coverage) tends to lower the retrieved [Fe/H]. Yet, the same is not the case when a synthetic library is used with its full HRD coverage. In the case of [Fe/H] inferred form Coelho14ext, we also see here the low-metallicity outliers discussed in Section 5.1.2. The systematic differences in inferred metallicities ($\Delta$[Fe/H]) are +0.05, +0.004, and +0.20 for Coelho14ext, SynCoMiL, and MILES, respectively.

### 5.4. *Old clusters: $\Delta t$ as a function of [Fe/H]*

Here we further investigate the effect seen in Fig. 9 in the regime of old ages. We show in Fig. 11 the results for the sub-sample of clusters with high-resolution spectroscopic metallicities obtained from literature (86 clusters, mostly from the Milky Way). The top panel (inspired by Figure 11 of

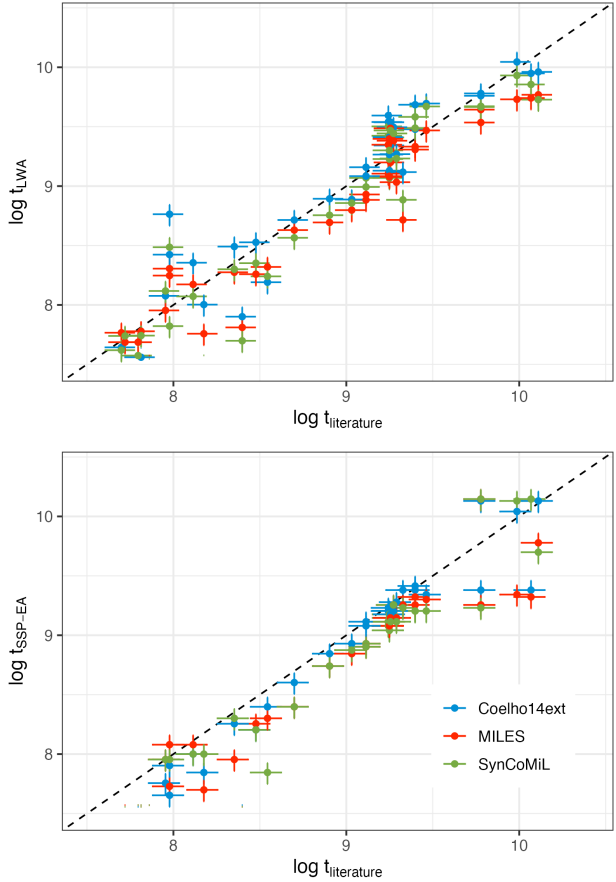

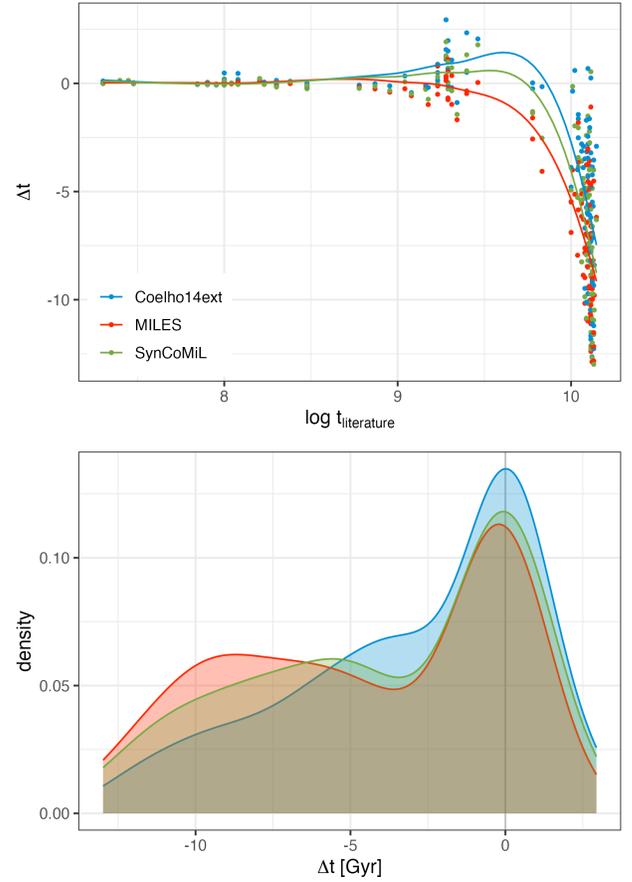

**Figure 8**: Comparison between the age of LMC and SMC clusters determined by Milone et al. (2023) (x-axis) and our STARLIGHT results (y-axis). The top panel shows LWA (Eq. 3) and the bottom panel the SSP-EA (Section 3).

**Figure 9**: Behaviour of $\Delta t$ (age derived from spectral fitting minus age from isochrone fitting) for our entire sample. Isochrone ages are from Milone et al. (2023) supplemented by the compilation in Gonçalves et al. (2020, Table 1). The top panel shows $\Delta t$ as a function of literature age. The tendency lines are obtained from LOESS. The bottom panel shows the density distribution of $\Delta t$. In both panels, the color code identifies the stellar library that was employed in modeling the SSP templates.

Gonçalves et al. 2020) shows $\Delta t$ as a function of metallicity, and the bottom panel shows the density distribution of $\Delta t$. By this comparison, we notice the same pattern previously pointed out in the literature: the ages of the Galactic metal-poor clusters are underestimated, in some cases by several Gyr, for all versions of the CBC20 models.

The same effect is seen in Gonçalves et al. (2020) using different SSP models. These authors used SSP models by Vazdekis et al. (2015), based on isochrones by Pietrinferni et al. (2006), and the stellar libraries MILES (Sánchez-Blázquez et al. 2006) and Coelho et al. (2005). We have in common only the MILES library in the case of our MILES-based models, although we adopt different stellar atmospheric parameters than the ones adopted by Vazdekis et al. (2015) (we refer the reader to Coelho et al. 2020 for a discussion on the atmospheric stellar parameters in MILES).

The classical explanation of the large $\Delta t$ value for old, low metallicity clusters, is the presence in these clusters of an extended horizontal branch, stars which are not included by default in the SSP templates (see, e.g., the recent discussion in Cabrera-Ziri & Conroy 2022). Additionally, a significant fraction of Blue Stragglers can also play a role in this context as they skew the integrated light toward younger ages, making age estimates in metal-poor systems more challenging. Many Galactic globular clusters (GCs) host multiple stellar populations with differing chemical abundances, including variations in helium and CNO elements (Bastian & Lardo 2018). While these variations have limited effects on isochrone age estimates in some optical bands (Cassisi & Salaris 2020), they may impact age and metallicity estimates from integrated light, although the extent is unclear. This complexity introduces uncertainties that may skew inferred ages and metallicities.

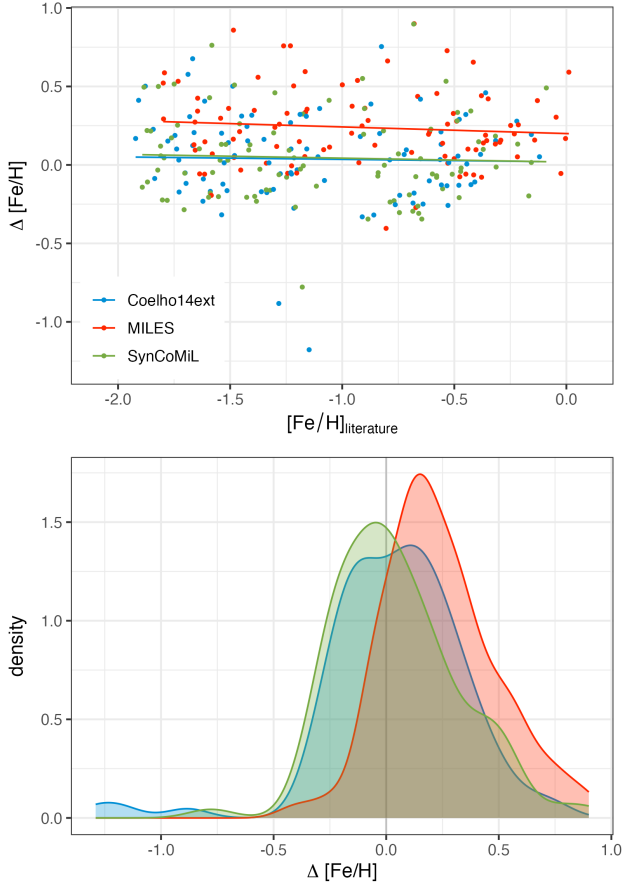

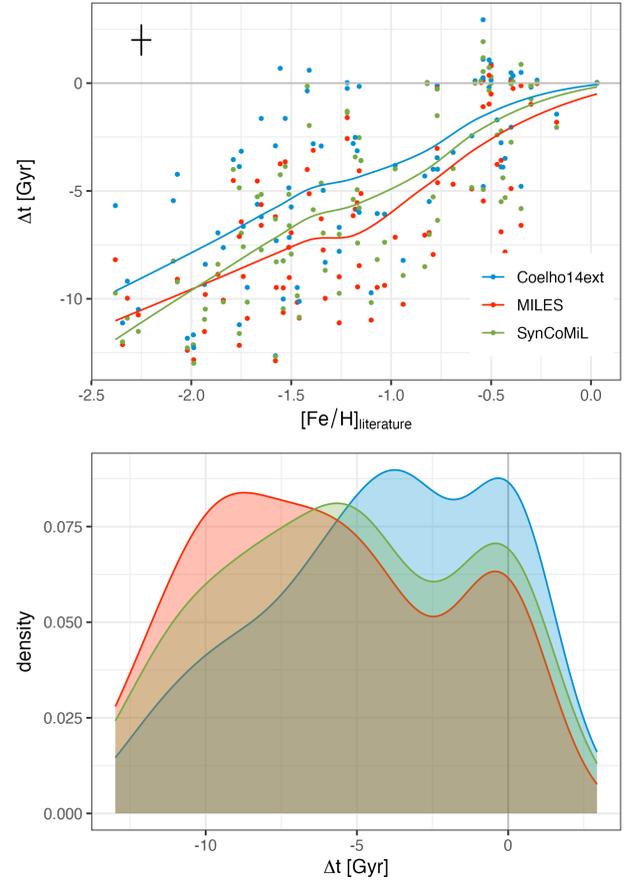

**Figure 10**: Top panel shows Δ[Fe/H] (Starlight minus literature). The color-coded lines are linear fits obtained for each SSP template. Bottom panel shows the density distribution of Δ[Fe/H] = [Fe/H]$_{\rm STARLIGHT}$ − [Fe/H]$_{\rm LITERATURE}$, color-coded by SSP template.

**Figure 11**: Behaviour of Δt (age derived from spectral fitting minus age from isochrone fitting) for the sub-sample of clusters with high-resolution metallicities from the literature. Isochrone ages are from Milone et al. (2023) supplemented by the compilation in Gonçalves et al. (2020, Table 1). The top panel shows Δt as a function of metallicity. The tendency lines are obtained from LOESS. The bottom panel shows the density distribution of Δt. In both panels, the color code identifies the stellar library that was employed in modeling the SSP templates.

The top panel of Figure 11 shows that the problem worsens for the MILES and SynCoMiL SPS models. This implies that the coverage effect can explain part of the observed behavior of Δt: when a better HRD coverage is available in the stellar library – Coelho14ext, in this case – Δt decreases for all [Fe/H]. The bottom panel of Figure 11 shows the density distribution of Δt, where clearly two peaks can be observed. The secondary peak, corresponding to the metal-poor clusters, peaks at around −3.5 Gyr for SSP models based on Coelho14ext, while it peaks at ≤ −6 Gyr for SSP models based on libraries with limited HRD coverage.

## 6. SUMMARY

We utilize a sample of 129 integrated spectra of star clusters of high SNR to examine the uncertainty associated with choosing an empirical versus a theoretical stellar spectral library in SPS models to use for full-spectral fitting techniques. Three SP models have been employed: one built with an empirical library, one built with a synthetic library with the same HRD coverage as the empirical one, and the last built with a synthetic library with optimal coverage of the HRD. These models allow us to separate the "synthetic effect" from the "coverage effect". The synthetic effect shows the changes introduced by replacing an empirical library with a theoretical one at fixed HRD coverage of stellar atmospheric parameters. The coverage effect shows the changes due to replacing the coverage of the HRD, between a sparse coverage typical of empirical libraries and an optimal coverage typical of synthetic libraries. We compare the results from fits using the different SPS models with values from the literature, namely, age from isochrone fitting and metallicity from high-resolution stellar spectroscopy.

From the model to model comparisons, these are our findings:

1. There is a good agreement between the metrics of the quality of fit when comparing the results obtained from the three models. The fits with SPS models built with synthetic libraries tend to have slightly higher $\chi^2$.

2. There is good agreement between the values of reddening obtained using the different SPS models.

3. Ages are affected by both the synthetic and the coverage effect. Models based on synthetic libraries tend to yield slightly lower ages. Models based on a sparse HRD coverage tend to overestimate the age of intermediate age populations.

4. The estimated [Fe/H] is slightly lower when using SPS models based on the theoretical SynCoMiL library than those based on the empirical MILES library. Most clusters with $\chi^2 > 3$ show up as outliers when fitted with Coelho14ext, resulting in very low inferred metallicity.

5. When limiting our fits to SSP equivalent parameters

   the results for age and reddening match well; however, the well-known "quantization" issue of metallicity when performing single SSP fitting is evident, due to the coarse coverage in [Fe/H] of the SPS model grids.

From the comparison of our spectral fitting results with parameters from the literature, we note that:

1. The distribution of $\Delta t$ has a distinctive double-peak pattern, for all SPS models: for clusters which are young or metal-rich, ages from integrated light are in agreement with literature; yet, all models underestimate the age of old and metal-poor systems, with models based on sparse HRD coverage being more affected than models with optimal HRD coverage. This second peak is around $-3.5$ Gyr for Coelho14ext models, and $\leq -6$ Gyr for SPS models based on libraries with limited HRD coverage.

2. All models predict metallicities which are slightly higher than values in the literature for metal-poor systems, and slightly lower than values in the literature for metal-rich systems.

3. On average, results from SPS models based on synthetic libraries match better the metallicity obtained from high-resolution stellar spectroscopy. The average differences in inferred metallicities ($\Delta$[Fe/H]) are +0.05, +0.004, and +0.20 for Coelho14ext, SynCoMiL, and MILES, respectively.

We would like to thank the anonymous referee whose comments greatly improved this manuscript. This work is supported by the FRG Grant and the Open Access Program from the American University of Sharjah. This paper represents the opinions of the authors and does not mean to represent the position or opinions of the American University of Sharjah. PC acknowledges support from Conselho Nacional de Desenvolvimento Científico e Tecnológico (CNPq) under grant 310555/2021-3 and from Fundação de Amparo à Pesquisa do Estado de São Paulo (FAPESP) process number 2021/08813-7. GB acknowledges financial support from the National Autonomous University of México (UNAM) through grants DGAPA/PAPIIT IG100319 and BG100622.